\documentclass[3p,times, procedia]{elsarticle}
\usepackage{ecrc}
\usepackage[bookmarks=false]{hyperref}
    \hypersetup{colorlinks,
      linkcolor=blue,
      citecolor=blue,
      urlcolor=blue}

\volume{00}

\journalname{Procedia Computer Science}

\runauth{M. Petrescu et al.}

\jid{procs}

\usepackage{amssymb}
\usepackage{subcaption}

\usepackage[figuresright]{rotating}

\begin{document}
\begin{frontmatter}



\dochead{27th International Conference on Knowledge-Based and Intelligent Information \& Engineering Systems (KES 2023)}%

\title{Students' interest in knowledge acquisition in Artificial Intelligence}



\author[a]{Manuela-Andreea Petrescu}
\author[a]{Emilia-Loredana Pop\corref{cor1}}
\author[a]{Tudor-Dan Mihoc}

\address[a]{Department of Computer Science, Faculty of Mathematics and Computer Science, Babe\c{s}-Bolyai University, 1 Mihail Kog\c{a}lniceanu street, 400084 Cluj-Napoca, Romania}
\email{manuela.petrescu@ubbcluj.ro, emilia.pop@ubbcluj.ro, tudor.mihoc@ubbcluj.ro}


\begin{abstract}
Some students' expectations and points of view related to the Artificial Intelligence course are explored and analyzed in this study. We anonymous collected answers from 58 undergraduate students out of 200 enrolled in the Computer Science specialization. The answers were analysed and interpreted using thematic analysis to find out their interests and attractive and unattractive aspects related to the Artificial Intelligence study topic.
We concluded that students are interested in Artificial Intelligence  due to its trendiness, applicability, their passion and interest in the subject, the potential for future growth, and high salaries. However, the students' expectations were mainly related to achieving medium knowledge in the Artificial Intelligence field, and men seem to be more interested in acquiring high-level skills than women. The most common part that wasn't enjoyed by the students was the mathematical aspect used in Artificial Intelligence. Some of them (a small group) were also aware of the Artificial Intelligence potential which could be used in an unethical manner for negative purposes. Our study also provides a short comparison to the Databases course, in which students were not that passionate or interested in achieving medium knowledge, their interest was related to DB usage and basic information. 
\end{abstract}

\cortext[cor1]{Corresponding author. Tel.: +40-745-085-130.}
\end{frontmatter}

\email{manuela.petrescu@ubbcluj.ro}



\section{{\uppercase{Introduction}}} 


Over the last decade, Artificial Intelligence (AI) has become a part of any individual's life, having an impact on almost all domains of activity and being seen as a machine's ability to mimic some people's capabilities, like learning, organizing, and reasoning. 

As AI models were developed, the requirement for data became increasingly pressing. The significantly higher need for information was not altered by the ongoing current disputes among computer scientists as to whether these models are actually discovering new information or only acquiring human knowledge from the labeled data sets, or by the one related to the complexity of the models versus the data sets size. Due to this necessity, over time, a significant relationship between artificial intelligence and databases naturally emerged.

An essential element in the development of AI expert models and in the development of knowledge-based systems is knowledge acquisition. By translating the human expertise into a proper machine representation, it serves as foundation for assessing, eliciting, and interpreting the information required for solving a specific problem. The employed techniques fall into three categories: manual, semi-automatic (interactive), and automated (machine learning and data mining) \cite{Ho2007}.  

Knowledge acquisition includes complex processes like recognition, learning, affiliation, and reasoning. Some of the specific activities, methodologies, and tools are presented in \cite{Jayashri2018}, as are examples of auto-generated database queries that have been created. At the core, the purpose of machine learning systems is to acquire knowledge. Real-world issues are tackled and examined from the perspective of gaining information \cite{Nedellec1995}. So the database procedures of knowledge discovery and acquisition are the process of analyzing vast quantities of information, uncovering new facts and expertise, and applying them to the resolution of business problems. This process was studied from the perspective of data mining in \cite{Bresic2012}, allowing a multidimensional data management and a graphical representation thereof. 

The human factor is one of the most important elements related to the impact of Artificial Intelligence and Databases (DB). For Mathematical and Computer Science or Technical and Engineering students, their socioeconomic status, motivation, performance, self-efficiency, and engagement in Computer Science are important factors that affect their knowledge and involvement in Artificial Intelligence and Databases \cite{Kahraman22,Spieler20}. These two  are of actual interest to the entire community, providing different perspectives and alternatives from the student's point of view and future aspects for our employees and employers. These domains are present in job requests on the labor market, and the aim is to find out if students want a career path related to AI and DB or if they just want to use them based on need. For example, in \cite{Petrescu2023C,Pop2023E}, the trend of replacing \textit{learning in-depth} with the more shallow \textit{get the work done} is presented.

To capture these aspects, we conducted an optional and anonymous survey (asking about the specialization and the gender) among the participants of an Artificial Intelligence course, from the Faculty of Mathematics and Computer Science, Babe\c{s}-Bolyai University, Cluj-Napoca, Romania. The participant set for this study was formed by students enrolled in the second year for Computer Science related specialization, and there were no constraints on their participation in this study or the types of answers they provided. 
The aim is to present the student's expectations related to Artificial Intelligence 
and their interest in following a career path in an AI-related domain by investigating the reasons behind their attractions and interests in AI. In a subsidiary, even if it was not the original scope of the paper, we make a comparison between the interest and desire to learn in men versus women. We had no hypothesis or assumptions related to results of the study; the approach was to define three research questions for which we found the answers using a survey (having multiple questions, the questions are detailed in \ref{sec:data_collection} section).
\begin{itemize}
    \item Are students interested in AI development / would they like to work in this field of computer science?
    \item What are the stated reasons for AI interest?
    \item Are any gender based differences in learning in depth interest?
\end{itemize}

The paper starts with an {\em Introduction} and a {\em Literature Review}, where studies related to knowledge acquisition in Artificial Intelligence and Databases are presented. It is followed by the {\em Methodology} section, where the survey setup is presented. The outcomes of Data collection and analysis are in the section titled {\em Results}. The {\em Threats to Validity} section summarizes the potential threats to validity and the actions taken to minimize and mitigate them; the {\em Conclusions and Future Work} section concludes the article.

\section{{\uppercase{Literature Review}}}

A wide range of career options, with applications in Databases and Artificial Intelligence (and not only), are accessible nowadays in the Computer Science domain. A lot of women that are pursuing employment are working in this field of activity with a great deal of enthusiasm, akin to the men active in this field of employment, who were the main type of employees in the previous years \cite{Kerkeley21,Spieler20,Weston19}.
When selecting a career path, it is important for both genders to prioritize efficiency, and a comparison should be made between them \cite{Mann20,OECD13}. These comparisons brought up aspects of professional interests that were not expected. For instance, a study conducted on youngsters aged between 13 and 19 revealed a professional track based on creativity. This was especially true for the women, who showed a tendency to become users of technology rather than makers of it \cite{Wong18}. 

The psychological, sociological, and developmental perspectives represent another source of guidance that is considered by the subjects when they choose the career path, as well as the individual needs for a sustainable future (external factors, internal improvements, path-related dynamism, and practical aspects) \cite{Hasan20}. Strengthening this idea is a survey about short-term and longer-term career aspirations and prior experience among the undergraduate IT students at an Australian university. The results indicate that the best motivation for a profession in these areas was the intrinsic interest in IT \cite{McKenzie22}.

Significant elements for the Databases field of activity were collected over the years from applied surveys of students and professors, but as far as we could tell, very little data was collected related to the social impact of modern AI models in the last few years. We have to consider the impact of these models in modern daily life (just take into account the Italian ban on chat GPT to understand their importance and influence).

This trend in society comes right after the COVID-19 global crises. To assess the impact of the changes made since that time, a survey was done in the United States one year after the breakout of the pandemic. It revealed that hybrid learning (online and onsite) was the optimal learning technique for college students \cite{Zhou21}.

Some divergent opinions held by college students concerning the possible consequences of AI were presented in \cite{Jeffrey2020}. They revealed how they felt about the future of AI development. Their knowledge and comprehension of AI were assessed, as were their faith in its potential advantages and their worries. A widespread worry about the technology's exponential growth and its potential impact on human employment was present even among those who see personal benefits from AI. Even if the respondents' knowledge of AI progress was high and so was the individual's belief in the advantages of AI, both skepticism and tension were obvious from the results. An old survey from Pega Systems in 2017 uncovered that while 72\% of respondents had some familiarity with Artificial Intelligence, 50\% were unfamiliar with the notion of AI as a technology that employs deep learning and natural language processing to solve complicated problems \cite{Pega2017}. The exponential increase in software implementation that attempts to collect the advantages offered by the novel capabilities of AI's inherent efficiency is having a profound effect on society \cite{Byrum2018}.

In a similar context, a gender comparison was presented from a comprehensive analysis of primary school teachers views regarding the impact of AI and its necessity in education \cite{Ryu2018}. The AI's judgments of female instructors proved to be less favorable than those of male teachers, and the value of education was deemed to be less important. Educators with expertise in top-tier schools understood that AI instruction would aid in fostering more creativity, and instructors with extensive teaching experience have a keen interest in AI and comprehend its importance. Another comparative study between the male and female students enrolled in the two major universities in Romania presented the variation in the amount of knowledge and interest in the field of Artificial Intelligence \cite{Gherhes2018}. The students pursuing technical courses proved to be more optimistic about the future of AI's sustainable development than the humanities students, who proved to be more concerned with defending human values and appeared more receptive to the drawbacks of AI advancement (economic crises, increase in military conflicts, negative effects on human relationships, fewer jobs for people).

Due to a study of perceptions and behavioral intentions toward learning AI performed in Beijing, was noticed that students will be better equipped for a future empowered by Artificial Intelligence if they have a higher purpose to study it \cite{Chai2021}.
With such a high impact on society and also on education it should be studied as early as possible, even in schools \cite{Seldon2018}. 
When confronted with a novel scenario, such as the rise of Artificial Intelligence, humans prefer to evaluate their choices. The result of such deliberation provides the cognitive basis for future decisions, so instructors should encourage self-efficacy and emphasize the benefits of AI \cite{Fishbein2015}. 

All these studies underline the importance of assessing the interest of students committed to an IT career in knowledge acquisition using artificial intelligence. 

\section{Methodology} 

The method applied in this study could be classified as survey research due to ACM Sigsoft Empirical Standards for Software Engineering Research \cite{ACM}. Our survey contains open and closed questions. The closed questions were used to find participants' gender.

The gathered information was used to measure and compare the parties intentions. Open questions, like {\it What do you like about AI? What are the reasons for which you find the field of AI attractive/unattractive?} were used to measure and have a better understanding of the students' intentions for in-depth learning. We had given one questionnaire to the students at the beginning of the course, and we had left it open for a couple of weeks to obtain relevant answers. Our interest was focused on the students, to measure their expectancy relative to AI, their perception of following a career path in AI-related fields, and their position on the subject. Their interest in AI can be discovered in the reasons that make AI attractive or unattractive. 

The participant set was formed by students from the Computer Science specialization enrolled in the Artificial Intelligence course.



\textit{Research Ethics}: Participation in the survey was anonymous and optional. When the students were asked to participate in the study, they were informed about the purpose of collecting the information and how it would be used. They also knew that the survey was anonymous and optional, and there were no other purposes except the ones mentioned.

\subsection{Participants}

Being an optional survey, only 58 students from 200 enrolled students participated in it and provided answers, 42\% of all the students. The participants were students enrolled in 
the second year of study. The survey was open for two weeks every time, thus allowing students to submit their responses. After two weeks we closed the survey, as we considered that the students who did not answer, probably don't want to participate in the study, so no other answers will be collected. We considered that in terms of percentage and numbers, the number of answers is comparable with the number of answers from other studies from the computer science domain \cite{marwan20,enase21}, thus making this study a valid one from this point of view. When we analyzed the obtained responses, we compared the percentage of the women that answered our survey to the total number of women enrolled in the course in order to validate that the women are correctly represented.

\subsection{Data Collection and Analysis}
\label{sec:data_collection}

As we wanted to have anonymous responses, we allowed students to answer at any time. We used Google Forms, a platform that students had previously used. The  survey contained closed questions to classify the received responses and open questions; the answers to the open questions were analyzed in depth. In terms of study classification, we used quantitative methods (questionnaire survey according to ACM community standards specifications \cite{ACM}). For interpreting the open question answers, we used thematic analysis \cite{Braun19}. These methods were previously used in other computer science-related studies {\cite{redmond13,easeai22,Kiger20}}. To obtain information, two authors analyzed the data by looking for specific keywords (items), then the items were grouped into classes. This process was validated by the other author (who did not participate in the process), and the last step was supposed to discuss different opinions and classifications for some items. Because some of the answers contained more than one item, the total number of items or keywords was greater than the total number of survey answers. The asked questions are mentioned in Table \ref{tab:AIquestions}.
\begin{table}[htbp!]
  \caption{\textbf{Survey questions}}
  \label{tab:AIquestions}
  {\small{
  \begin{tabular}{p{16.0cm}}
\toprule
   Q1. Gender (Choice of \textit{Male / Female / I don't want to answer})\\ 
   Q2. What are your expectations related to Artificial Intelligence course? \\
   Q3. What specifically would interest you regarding the field of Artificial Intelligence? \\
   Q4. What are the reasons for which you find the field of Artificial Intelligence attractive?\\
   Q5. What are the reasons for which you DO NOT find the field of Artificial Intelligence attractive?\\
   \toprule
\end{tabular}
}}
\end{table}

\section{{\uppercase{Results}}}

\subsection{What were your expectations related to the Artificial Intelligence course?}

\begin{figure}[htbp!]
\begin{subfigure}{0.5\textwidth}

    \includegraphics[width=1\linewidth ]{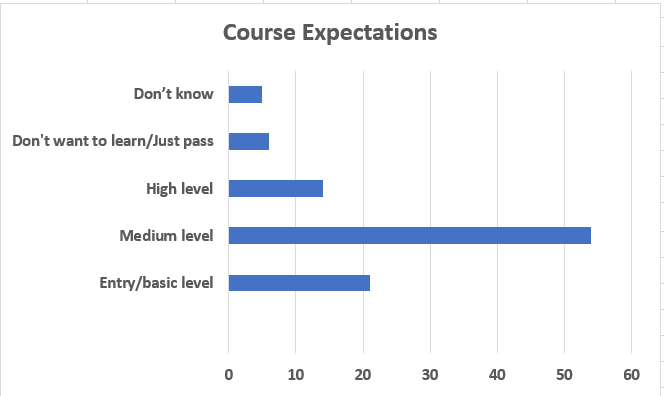} 

\end{subfigure}
\begin{subfigure}{0.5\textwidth}
   
\includegraphics[width=1\linewidth]{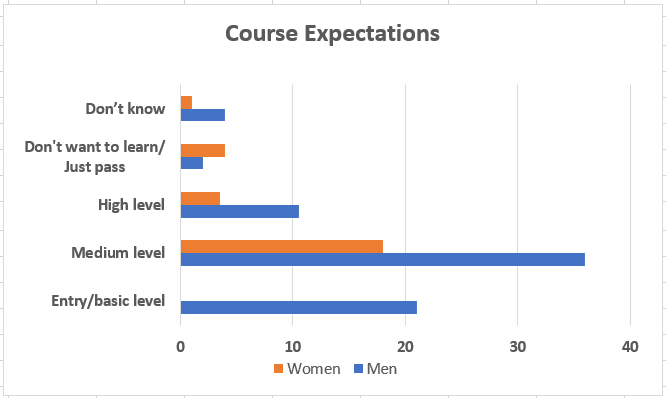}

\end{subfigure}
\caption{(a) Course Expectations ; (b) Course Expectations - Gender Based comparison.}
\label{fig_1}
\end{figure}

The Artificial Intelligence course was an introductory one, so their expectations may vary based on their prior knowledge and interests. Out of the surveyed group, 21\% students identified themselves as wanting to learn at an entry-level or basic level, indicating a desire for foundational knowledge and skills. Another 54\% students stated that they want to learn at a medium level, suggesting a desire to deepen their knowledge and skills beyond the basics. Only 14\% students stated they want to learn advanced topics, 6\% stated they just want to pass or don't want to learn; and 5\% stated that they don't know. In terms of gender comparison, women don't seem as eager to learn advanced/high-level skills (only 3.5\% women compared to 10.5\% men), they state they want to learn medium-level skills in the same percent as men do (30\% of the total number of women or 18\% of the total number of students). A difference appears in the \textit{''Just want to pass''} class, where the percentage of women is higher compared to the percentage of men who stated this option. The numbers (expressed in percentage) for students' expectations and also men vs. women comparison can be visualized in Fig. \ref{fig_1}.


The student's answers could be summarized in \textit{''application of basic knowledge in Machine Learning (supervised, unsupervised, reinforcement learning, neural networks, and decision trees) + (super nice to have) competitions on Kaggle'', ''I hope to better understand the mechanism behind programs and to be able to apply it myself or in other projects'', ''I can't say that I'm passionate about the field or that I want to delve into it in the future.'', 'I just want to pass the exam''}. 

\textbf{In conclusion}, students' expectations were related mainly to achieving medium knowledge in the AI-related field, and the men were a little more interested in acquiring high-level skills.\newline

\textbf{Students expectations comparison: Artificial Intelligence versus Databases course}

Similar studies \cite{Petrescu2023C,Pop2023E} have analyzed the student's expectations related to a database course previously studied and performed by the similar group of students from the same university in a similar context (introduction course, second-year students at the Computer Science specialization). The conclusion was that students were interested in achieving basic knowledge in the database-related field and SQL statements; they wanted to be able to create or use the DB knowledge at work or in personal projects but did not show a lot of interest in acquiring more in-depth knowledge.
The students seemed to be more interested in the AI course compared to the DB course; the main mentioned reasons were the hype/interesting domain and the domain's potential (to grow and to earn more as a developer).

\subsection{What specifically would interest you regarding the field of Artificial Intelligence?}

Interest in a specific domain and passion are closely correlated with knowledge acquisition, as passion and interest can serve as powerful motivators for learning. A passionate individual is more likely to be curious and driven to learn. This curiosity can lead to a deeper level of engagement and immersion in the subject matter, which in turn can enhance their ability to acquire new knowledge and skills. Moreover, interest and passion can help individuals overcome obstacles and setbacks that often accompany the learning process, leading to more successful outcomes in knowledge acquisition. Thus, our interest in the student's passion is legitimate with the scope of the paper, as their answers predict a higher level of knowledge acquisition.

\begin{figure}[htbp!]
\begin{subfigure}{0.5\textwidth}
 
    \includegraphics[width=1\linewidth ]{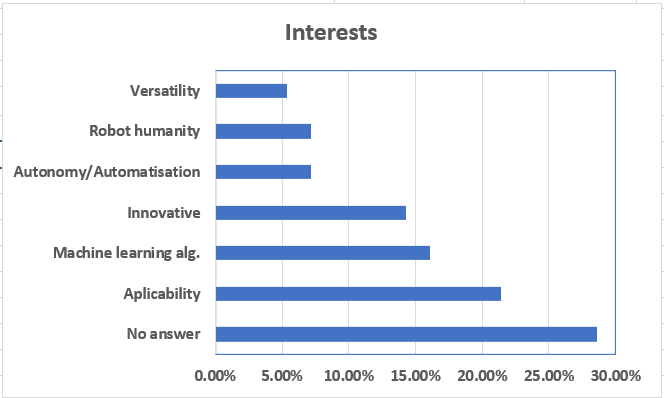} 
\end{subfigure}
\begin{subfigure}{0.5\textwidth}

\includegraphics[width=1\linewidth]{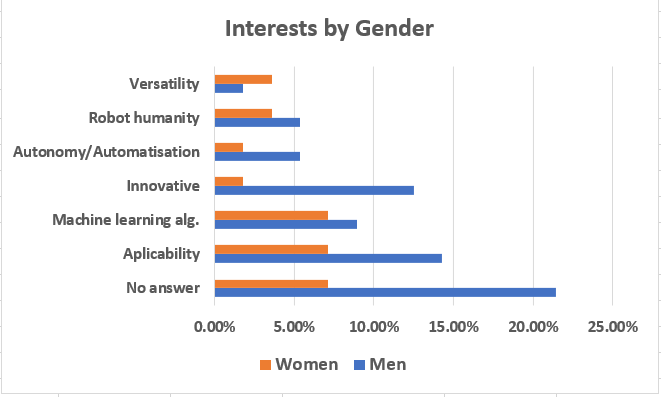}
\end{subfigure}
\caption{(a) Students' interest regarding the field of Artificial Intelligence; (b) Gender-based students' interest regarding the field of Artificial Intelligence.}
\label{fig_2}
\end{figure}

The fact that a significant percentage of students (28.57\% from which 21.43\% men and 7.14\% women) have chosen not to answer \textit{''Don't know''} regarding their interests in an AI course may reflect a reluctance to commit to a specific area of study, so they are not particularly interested in AI. However, the second larger group of students, 21.43\% expressed an interest in applicability, compared to 16.07\% in machine learning algorithms and 14.29\% in innovation. An interest in applicability suggests a focus on practical skills and knowledge that can be directly applied to real-world problems and challenges in a variety of domains: \textit{''The most exciting part is definitely automatization, or autonomous vehicles.'', ''AI in medicine'', ''Competetive AI for games''}. The stated interest in machine learning and algorithms suggests an interest in profound learning and more knowledge acquisition. A small group mentioned they are interested in AI's autonomy and automatization processes 5,36\%, and there were answers related to \textit{''Robot Humanity''}: \textit{''The fact that we can teach machines how to behave like humans''}. The percentages, total and gender-based, can be visualized in Fig. \ref{fig_2}.


Taking into account that the number of men was approximately three times the number of women, the only significant gender-based difference represented by stated interests is represented by the interest in innovation (where women seem less interested) and in machine learning (where women seem more interested compared to men).
\newline

\textbf{Students interests comparison: Artificial Intelligence versus Databases course}

For the AI course, we could not say for sure that men are more interested in learning in-depth compared to women. They are more interested in autonomy/automatization, in the innovation and applicability parts of AI. Women declared themselves (in the largest percent) interested in Machine Learning algorithms (the values are comparative with men's values). But if we take into consideration that women are three times smaller than men, the major difference is the innovative part. We could not establish a correlation between gender and terms such as security, and system integration, as these terms did not appear in their answers. The terms used for expressing their interests are correlated to course specificity: in a DB course survey \cite{Petrescu2023C,Pop2023E} (similar participant set) their stated interests were related to security, complexity, and efficiency. The study  conclusion \cite{Petrescu2023C} was that men were more interested in learning advanced features related to DB. This conclusion (men are more interested in learning advanced features) is sustained in our actual study by men's interest in the innovative aspect and in autonomy/automatization, topics in which fewer women stated they have an interest.

\subsection{What are the reasons for which the students find / do NOT find the field of Artificial Intelligence attractive?}

In terms of number, the primary reason why students are drawn to AI is because it is currently a \textbf{popular trend}, 35.71\% of students mentioning this element: \textit{''hype'', ''it's interesting'', ''New-ish field, many opportunities, futuristic'', ''It't in trend, and it seems to have an active community."}. In recent years, AI has been at the forefront of technological advancement in medicine and the automotive industry. This trendiness, \textit{''hype''}, as it was named by students, can be attractive due to its novelty and the fact \textit{''AI did not reach its full potential''}. \textbf{Passion} or the fact they are interested in the AI domain, was the second largest reason for attractiveness, mentioned by 30.40\% male and 5.36\% female students. \textbf{ The potential for future growth} as a reason for attractiveness is also mentioned by 14.30\% male and 12.5\% female students. As technology advances, AI is expected to play an increasingly important role in many industries, such as domain applicability, which was mentioned by 17.85\% of the students (mostly men). This means that there will likely be a high demand for AI experts in the coming years, making it a potentially lucrative career choice. The potential for \textbf{high salaries} is another attractive feature; 7.10\% mentioning the \textit{''money'', ''it's paid above average''} as a reason for domain attractiveness. Only 12,5\% of the students did not answer this question, so the above conclusions were considered relevant. \newline

\textbf{Reasons for which the students do NOT find the field of Artificial Intelligence attractive?}

\textbf{Mathematics} is a fundamental subject; however, not all students enjoy studying or using math, not even those in Computer Science. To our surprise, mathematics was the number one reason (33\% of the answers) mentioned in the students' answers for which they don't like AI. Some answers were short and concise: \textit{''mathematics''}, others contained some details: \textit{''very complex, low-level, math-heavy field'', ''you need advanced mathematical knowledge'', ''mathematics is not my strong point''} or \textit{''A lot of maths and high possibility of failure''}. Another reason mentioned by the students was the subject's \textbf{difficulty}, difficult concepts, and a lot of complexity overall. Around 14.28\% of the answers mentioned this aspect: \textit{''pretty difficult concepts'','its difficulty'', ''Much too abstract.'', ''Statistics and the mathematics models complexity.''}.

A surprising reason that appeared in 14.28\% answers was related to AI's potential to be used for \textbf{negative purposes}: \textit{''The fact that a powerful tool in the wrong hands can lead to many ugly scenarios...'', ''The lack of ethics that can be encountered in the field''} or \textit{''I don't really like things that try to resemble humans, like Sophia or chatbots''}. Some answers mentioned and took into consideration the economic impact in long term: \textit{''Big tech companies will have very powerful processing capabilities, a lot of data, and it will be very difficult to compete with them''}. Additionally, some students mentioned that the AI field is \textbf{over-hyped}, 7.14\% mentioning the overall attention and hype that AI is creating in the society: \textit{''I believe there is more hype than what can currently be achieved.'', ''There is too much hype about AI at the moment''}. A small number of answers, 5.35\% of the total number of answers, stated the fact that they prefer other courses/subjects, mentioned by the \textbf{large volume of data} and the difficulty to find relevant data to train the models (3.57\% of answers) \textit{''The necessity of a large volume of data to produce significant progress''}. 17.85\% of the students did not provide any answer to this question.

In conclusion, students are considering AI for a variety of reasons, including its trendiness, their passion and interest in the subject, the potential for future growth and applicability, and the potential for high salaries. Mathematics seems to be the number one reason why they don't like AI, and surprisingly, the students seem to be aware of AI's potential to be used in a non-ethical manner, for negative purposes.

\section{{\uppercase{Threats to Validity}}} 

For the threats to validity, we tried to mitigate them according to the recommendations described in \cite{ACM}. Our survey analysis was focused on the participants' selection, target participant set, contingency actions for drop-outs, and research ethics. The target participation set was formed by the second-year students enrolled in the Artificial Intelligence and Databases courses, on the specialization {\em Computer Science} in the English line of study. The students were organized into study groups, each having up to 30 students, alphabetically ordered by their surname. The groups were randomly allocated to a teacher, and each student from the group could choose to take part or not in the survey. In conclusion, in terms of the data set, the student allocation was completely random, thus ensuring a representative group of participants. In terms of number and target group representation, we collected answers from 58 of the 200 enrolled students, or around 29\% of the total number of students. As the survey was optional, we could not enforce a larger student's participation, but as previously mentioned, other studies in computer science have a comparable number of participants. Because of this, we considered that our study meets the sufficient participation criteria.
The questions were optional, so the students could skip any question they did not want to answer. The collected data was analyzed using thematic analysis, and we followed the recommendations to mitigate the risk of a subjective approach.

As for research ethics, as we mentioned before, the students were informed about the purpose and usage of the collected data before participating in the survey.

\section{{\uppercase{Conclusion and Future Work}}} 

AI has gained a lot of exposure in the last few years due to the progress of its applications in various domains. In this context, the scope of the paper was to find out if the students in Computer Science, second year, were interested in AI and what they liked or disliked about AI. On specific topics, we made a comparison with the results obtained in a previous study of DB interests for a similar set of participants. We had an anonymous and optional survey where students stated their interest in learning AI, with most of them stating they want to learn at the medium level, the second largest group stating they want to learn the basics, the third group wanting to learn at the high level, and the smallest group not answering or mentioning that they had not decided. In the previous DB study, most of the students stated that they wanted to learn the basics, so their expectations related to the AI course are a little higher. When asked in detail, the students' interests lay between applicability, machine learning algorithms, innovative aspects, automatization, and robot humanity. It was interesting that almost 30\% did not answer this question, and men scored higher in interest in applicability and innovative aspects compared to women. According to the data from \cite{Petrescu2023C, Pop2023E}, the interest for in-depth learning in databases was more visible in men compared to women: they wanted to learn about \textit{''security'', ''efficiency'', ''system integration''}; fewer women mentioned such interests.The innovative aspect, or \textit{''the hype''}, generated by AI was mentioned in many answers, mainly as a reason for liking AI, but we had a few answers where the hype was considered negative: \textit{''too much hype''}. In terms of applicability, mathematics and complexity were the two major reasons mentioned by the students for not liking AI. Another aspect mentioned was the possibility of using AI technology for malicious purposes.

A prevalent reason that did not appear in the previous study was the reason related to \textit{''money''}. The AI domain is perceived as offering opportunities for growth and for bigger salaries. It would be worth investigating whether the higher interest in AI compared to DB is because of the financial aspects, the \textbf{''hype''}, or a combination of both. To find out an answer, a similar study can be performed for other courses to compare the students' interests and motivation.

\end{document}